\documentclass[12pt]{article}
\usepackage{amsmath}
\usepackage{graphicx}
\usepackage{enumerate}
\usepackage{natbib}
\usepackage{bm}
\usepackage{url} % not crucial - just used below for the URL
\usepackage{booktabs}
\usepackage{amssymb,amsfonts,amsthm}
\usepackage{tikz}
\usetikzlibrary{positioning}
\usetikzlibrary{arrows,arrows.meta}
\RequirePackage[colorlinks,citecolor=blue,urlcolor=blue]{hyperref}

\hypersetup{backref, final=true, pdfpagemode=FullScreen,  colorlinks=true}

\usepackage{amssymb}
\usepackage{longtable}
\usepackage{lineno,hyperref}
\usepackage[section]{placeins}
\usepackage{multirow}
\usepackage{array}
\usepackage{algorithm}
\usepackage{algorithmic}

\makeatletter
\def\singlespace{\def\baselinestretch{0.7}\@normalsize}
\def\beqn{\begin{eqnarray}}
	\def\eeqn{\end{eqnarray}}
\def\beqns{\begin{eqnarray*}}
	\def\eeqns{\end{eqnarray*}}
\def\sgn{\mbox{\rm sgn}}

\newtheorem{Remark}{Remark}%[section]
\newtheorem{theorem}{Theorem}%[section]
\newtheorem{Example}{Example}%[section]
\newtheorem{Proposition}{Proposition}%[section]

\newcounter{thm}

\newcommand{\bldsi}[1]{\mbox{\tiny  $#1$}}
\newcommand{\bldsr}[1]{\mbox{\footnotesize  $#1$}}

\def\sgn{\mbox{\rm sgn}}

\def\bbeta{\mbox{\boldmath{$\beta$}}}

\def\bPhi{\mbox{\boldmath{$\Phi$}}}

\def\bmu{\mbox{\boldmath{$\mu$}}}

\def\brho{\mbox{\boldmath{$\rho$}}}
\def\diag{\mbox{\rm diag}}

\def\bW{\mathbf W}
\def\bX{\mathbf X}

\def\bA{\mathbf A}
\def\ba{\mathbf a}

\def\bC{\mathbf C}
\def\bD{\mathbf D}

\def\bE{\mathbf E}
\def\be{\mathbf e}
\def\mF{\mathbf F}
\def\bG{\mathbf G}
\def\bH{\mathbf H}

\def\bQ{\mathbf Q}

\def\bV{\mathbf V}
\def\bW{\mathbf W}
\def\bY{\mathbf Y}
\def\bb{\mathbf b}

\def\mf{\mathbf f}

\def\bv{\mathbf v}

\def\b0{\mathbf 0}

\def\meta{\mbox{\boldmath{$\eta$}}}
\def\bbeta{\mbox{\boldmath{$\beta$}}}

\def\bmu{\mbox{\boldmath{$\mu$}}}

\def\brho{\mbox{\boldmath{$\rho$}}}

\def\diag{\mbox{\rm diag}}

\def\Var{\mbox{\rm Var}}

\def\E{\mbox{\rm E}}

\begin{document}
		\def\spacingset#1{\renewcommand{\baselinestretch}%
		{#1}\small\normalsize} \spacingset{1}
	
	\title{Penalized weighted GEEs for high-dimensional longitudinal data with informative cluter size}        % Enter your title between curly braces
	%\author{Yue, Ma}
	\author{\hspace{0.7cm}Yue Ma$^1$, Haofeng Wang$^2$ and Xuejun Jiang$^{1,}$\footnote{Corresponding author}     \\ 
		\vspace{0.1cm}
		\hspace{ 1cm}\textsl{ $^1$Department of Statistics and Data Science, Southern University  
			} \\
		\hspace{ 0.1cm} \textsl{of Science  and Technology, China}\\
		\vspace{0.1cm}
		\hspace{ 0.2cm}\textsl{ $^2$Department of Mathematics, Hong Kong Baptist University,
		} \\
		\hspace{ 0.2cm} \textsl{Hong Kong}
		}

	\date{}      % Enter your date or \today between curly braces
	\maketitle

	\noindent{\bf Abstract:}
    High-dimensional longitudinal data have become increasingly prevalent in recent studies, and penalized generalized estimating equations (GEEs) are often used to model such data. However,  the desirable properties of the GEE method can break down when the outcome of interest is associated with cluster size, a phenomenon known as informative cluster size. In this article, we address this issue by formulating the effect of informative cluster size and proposing a novel weighted GEE approach to mitigate its  impact, while extending the penalized version for high-dimensional settings. We show that the penalized weighted GEE approach achieves consistency in both model selection and estimation. Theoretically, we establish that the proposed penalized weighted GEE estimator is asymptotically equivalent to the oracle estimator,  assuming the true model is known. This result indicates that the penalized weighted GEE approach retains the excellent properties of the GEE method and is robust to informative cluster sizes, thereby extending its applicability to highly complex situations.
    Simulations and a real data application further demonstrate that the penalized weighted GEE outperforms the existing alternative methods.

	~\\
	

\newpage
\spacingset{1.9} % DON'T change the spacing!
\section{Introduction}
\label{s:intro}

Longitudinal data, characterized by repeated observations of correlated responses across a wide range of covariates, frequently occur in fields such as biology, social sciences and economics. \cite{Manuel} provides extensive modelling of longitudinal data in economics, documenting factors such as cost, production, demand and profit across various regions over several decades. Recently, the generalized estimating equation (GEE) approach \citep{Liang,Wang2011} has gained widespread use due to its robust properties: (i) GEE  is optimal when the working correlation structure is correctly specified, and (ii)
it is robust to misspecified correlation structures.

Despite the robustness and optimality of the GEE approach, the GEE estimator becomes biased when the cluster size is informative (Proposition \ref{prop1}). In practice, it is common to encounter scenarios where the cluster size is nonignorable. For example, patients in good condition after recovery may not seek medical attention again. Similarly, in the periodontal study conducted by \cite{GWM1998}, individuals with poor dental health were found to have fewer teeth than their counterparts. Different cluster sizes can either be viewed as data missing (as in the examples  above) or as a feature generated from the data itself \citep{SPC2014}. There are two kinds of clustered data analysis in the literature. The first one is complete-cluster inference, where missingness is random and the response variable $Y$ is independent of the cluster size $M$.  The other one is called observed-cluster inference, where the loss of observations conveys some information about the cluster, meaning that $Y$ depends on $M$.  This phenomenon is termed as informative cluster size (ICS), also referred to as nonignorable cluster size \citep{Hoffman2001,SPC2014}. %WDS003,PSC2013,SPC2014}.
In the aforementioned periodontal study \cite{GWM1998}, individuals with poor dental health were found to lose more teeth than those with good dental health, leading to small cluster sizes. Meanwhile, their risk of dental disease is higher than that in healthy individuals. In a toxicology experiment \citep{Hoffman2001}, pregnant dams were randomly exposed to a toxicant. Dams that are susceptible to the toxicant may produce pups with a high rate of birth defects and experience more foetal resorption than their counterparts, resulting in small cluster sizes.

In this article, we focus on observed-cluster analysis, where the mean of response is related to the cluster size. To address the challenge posed by ICS, we first analyze the expectation of GEEs (Proposition \ref{prop1}) and gain insight into  the effect of ICS. This motivates us to propose the weighted GEE (WGEE) for reducing the bias and recovering the consistency of GEE estimators (Section \S \ref{Sec2.2}). For high dimensional settings, we propose  the penalized WGEE (PWGEE) approach to simultaneously perform variable selection and estimation  in the presence of ICS.
Theorem \ref{Th1}  establishes  model selection consistency and estimation consistency, while
Theorem \ref{Th2} demonstrates the asymptotic normality of the PWGEE estimator.
These theorems show that our WGEE approach retained the desirable properties of the GEE method and was robust to ICS, thereby extending the applicability of GEE to highly complex situations.

Some estimation and inference results are available for high dimensional longitudinal data.
\cite{Wang2009} introduced a BIC-type model selection criterion based on the quadratic inference function. \cite{Wang2012} used the penalized GEE (PGEE) method to perform variable selection and estimation simultaneously when the first two marginal moments and a working correlation structure are pre-specified. \cite{FNL20} proposed a quadratic decorrelated inference function approach to make inferences for high-dimensional longitudinal data. \cite{XS2023} constructed a one-step de-biased estimator via projected estimating equations for GEE to test significance on any linear combinations of high-dimensional regression coefficients in longitudinal data.
However, the aforementioned methods are invalid under scenario in which the cluster size is informative.
Moreover, researchers adopted resampling methods to address the effect of ICS. \cite{Hoffman2001} proposed the within-cluster resampling (WCR) method, a computationally intensive Monte Carlo procedure, to estimate regression coefficient under marginal models. \cite{RW2002} investigated the within-cluster paired resampling method to estimate a cluster-specific exposure parameter for analyzing clustered binary outcome data.
%\cite{WDS003} proposed cluster-weighted GEE  to fit marginal models for longitudinal data, avoiding the computationally intensive resampling procedure in WCR.
%Similarly, \cite{BRS2005} investigated the inverse cluster size-weighted estimating equation to obtain asymptotically valid inferences when ICS is present.
\cite{CL2008} proposed the modified WCR  approach to increase estimation efficiency, but it requires the minimum cluster size to be greater than one and imposes additional restrictions on the correlation structure. \cite{SC2018} developed a model selection criterion called the resampling cluster information criterion for semiparametric marginal mean regression in finite dimensions. However, the above WCR methods rely on the average of plentiful resampled-based estimators. When $Q$ resamplings are conducted,	the WCR estimator is calculated as $Q^{-1}\sum_{q=1}^{Q}\hat\bbeta^{(q)}$, where $\hat\bbeta^{(q)}$ is obtained from the $q$th resampled data set for $q=1,\ldots,Q$. It's well known that there is no theoretical guarantee in model selection for the  averaging method in high-dimensional regression, as it may lead to overfitting \citep{Wel23}.

In this article, we propose the PWGEE method for  simultaneous variable selection and estimation in high-dimensional longitudinal data with the presence of ICS. Our PWGEE approach specifies only the first two marginal moments and a working correlation matrix, thereby avoiding the complexity of fully specifying a joint likelihood when analyzing  correlated data. Compared with the existing approaches, the major contributions of this paper are as follows.
First, we formulate the problem of ICS and illustrate the classical GEE estimator is biased with ICS (Proposition \ref{prop1}). Second, we proposed the WGEE method by introducing weights into the generalized estimating equations %the working correlation matrix of responses
to eliminate the influence of ICS. The WGEE method is robust to the choice of working correlation matrix whether the cluster size is informative or not.
On this basis, we further propose the PWGEE for simultaneous variable selection and estimation in high-dimensional longitudinal data.
Third, we establish model selection consistency and asymptotic normality of the PWGEE estimator, an oracle property in model selection. In addition, we introduce a fast iterative algorithm for the PWGEE that shrinks the coefficients of redundant variables to zero, leading to a sparse solution. Last, our PWGEE approach retains the advantages  of the GEE method while remaining robust to ICS, thereby extending its applicability to highly complex scenarios.

This article is organized as follows. In Section \ref{Sec2}, we formulate the problem of ICS, introduce the WGEE method in detail and  describe the PWGEE algorithm. In Section \ref{Sec3}, we establish the theoretical properties of the PWGEE method. In Section \ref{Sec4}, we conduct simulation studies and present performance comparisons. In Section \ref{Sec5}, we apply the methods proposed to plasma proteome expression data from  patients with COVID-19. In Section \ref{Sec6}, we provide a brief discussion. The proofs of the theorems are delegated into the Appendix. %Supplementary Materials.

Throughout this article, we use the following notations. For any constants $a$, $[a]$ denotes its integer part. For vectors $\ba=(a_{1},\ldots,a_{m})^T\in \mathbb{R}^m$ and $\bb=(b_{1},\ldots,b_{m})^T\in \mathbb{R}^m$, we define $|\ba|=(|a_1|,\ldots,|a_m|)^T$, $\Vert\ba\Vert_{1}=\sum_{i=1}^{m}|a_{i}|$, $\Vert \ba\Vert_{2}=\sqrt{\sum_{i=1}^{m}a_{i}^2}$, $|\ba|_{\infty}=\max_{1\leq i\leq m}|a_i|$, and $\ba\circ\bb=(a_1b_1,\ldots,a_m$
$b_m)^T$. For any subset $\mathcal{M}$ of the row index set  $\{1,\ldots,m\}$ of $\ba$,  $|\mathcal{M}|$ denotes the cardinality of $\mathcal{M}$, and $\mathbf{a}_{\bldsi{\mathcal{M}}}$ represents the subvector of $\ba$ corresponding to the indices in $\mathcal{M}$. The indicator function is denoted by $\mathbb{I}(\cdot)$.

Let $\bE$ be a  $m\times m$ matrix  and $\bE^{-1}$ be the inverse of $\bE$. For $\mathcal{M}_1,\mathcal{M}_2\subseteq \{1,\ldots,m\}$, $\bE_{\mathcal{M}_1}$ represents the submatrix of $\bE$ formed by column indices in $\mathcal{M}_{1}$, and $\bE_{\mathcal{M}_{1},\mathcal{M}_{2}}$ represents the submatrix whose row and column  are indexed by $\mathcal{M}_{1}$ and $\mathcal{M}_{2}$, respectively. The maximum and minimum eigenvalues are denoted by $\lambda_{\max}(\cdot)$ and $\lambda_{\min}(\cdot)$, respectively. Finally, $\|\bE\|_{2}=\sqrt{\lambda_{\max}(\bE^T\bE)}$.

\section{Methodology}\label{Sec2}
\subsection{Problem Formulation}\label{Sec2.1}

\hspace{3mm} Let $Y_{i j}$ represent the response variable of the $j$th observation in the $i$th cluster, with a $p_n$-dimensional covariate vector  $\mathbf{X}_{ij}=(X_{ij,1},\dots,X_{ij,p_n})^T$, where $i=1,\dots,n$, $j=1,\dots,M_i$, and $M_i$ is referred to cluster size. We consider the first two marginal moments of $Y_{i j}$ as follows:
\begin{equation}\label{MT1}
	\E(Y_{ij}\mid \bX_{ij})=\mu(\bX_{ij}^T\bbeta^*) \ \ \text{and} \ \
	\Var(Y_{ij}\mid \bX_{ij})=\phi(\bX_{ij}^T\bbeta^*),
\end{equation}
where $\bbeta^*$ is an unknown $\mathbb{R}^{p_n}$ vector,
$\mu$ is a known link function and $\phi$ is the working variance.
It's worth noting that we allow for the possibility that $\phi$ may be misspecified. Our proposed methods do not require $\phi$ to represent the true variance function. Observations within each cluster are correlated,  i.e. $|{\rm Corr}(Y_{ij}, Y_{ik})|>0$, but observations are independent across clusters.
%The existing popular methods for the longitudinal data are mainly constructed under the framework of generalized estimating equations (GEEs)\citep{Liang}.
Let   $\bY_{i}= (Y_{i1},Y_{i2},\dots,Y_{iM_i})^T$ and $\bX_i=(\bX_{i1},\bX_{i2},\dots,\bX_{iM_i})^T$.

Define $\bmu_i(\bbeta)=(\mu(\bX_{i1}^T\bbeta),\ldots,$
$\mu(\bX_{iM_i}^T\bbeta))^T$
and $\bPhi_i(\bbeta)=\diag(\phi(\bX_{i1}^T\bbeta),$
$\ldots,\phi(\bX_{iM_i}^T\bbeta))$.
The optimal estimating equations for $\bbeta^*$ is given by
\begin{equation}\label{MT2}
	n^{-1}\sum_{i=1}^n \bX_i^T\bmu_i'(\bbeta) \bV_i^{-1}
	(\bY_{i}-\bmu_i(\bbeta))=\b0,
\end{equation}
where $\bV_i$ is the covariance matrix of $\bY_i$, and $\bmu_i'(\bbeta)=\diag(\mu'(\bX_{i1}^T\bbeta),\ldots,$$\mu'(\bX_{iM_i}^T\bbeta))$,
with $\mu'(\cdot)$ being the derivative of $\mu(\cdot)$. In practice, the  covariance structure of $\bV_i$ is often unknown. The commonly used approach is to introduce a working correlation matrix $\bG_i$ such that $\bV_i=\bPhi_i^{1/2}(\bbeta)\bG_i\bPhi_i^{1/2}(\bbeta)$. Let $\mF_i(\bbeta)=\bmu_i'(\bbeta)\bPhi_i^{-1/2}(\bbeta)$. Substituting  $\bV_i$ into Equation  \eqref{MT2} reduces it to
\begin{equation}\label{MT3}
	n^{-1}\sum_{i=1}^n\meta_i(\bbeta)=\b0,
\end{equation}
where  $\meta_i(\bbeta)=\bX_i^T\mF_i(\bbeta)\bG_i^{-1}\bPhi_i^{-1/2}(\bbeta)(\bY_{i}-\bmu_i(\bbeta))$ is a $p_n$-dimensional vector.

The GEE method is widely used in longitudinal data analysis. However, one key assumption underlying the GEE method is that outcomes are independent of the cluster size $M_i$'s. This assumption may not hold when outcomes are influenced by the cluster size $M_i$'s, leading to $E(Y_{ij}\mid \bX_{ij}, M_i)\not=E(Y_{ij}\mid \bX_{ij})$, a scenario referred to as ICS \citep{Hoffman2001}. %,WDS003,PSC2013,SPC2014}.
The following proposition illustrates that the GEE method results in biased estimates when the cluster size is informative.

\begin{Proposition}\label{prop1}
Let $\mf_{kl}(M_i)=E\{\bX_{ik}\mu'(\bX_{ik}^T\bbeta^*)\phi^{-1/2}(\bX_{ik}^T\bbeta^*)
\phi^{-1/2}(\bX_{il}^T\bbeta^*)(Y_{il}-\mu(\bX_{il}^T\bbeta^*)) \mid M_i
\}$, $k,l$
$=1,\ldots,M_i$. Denote by $\bG_i^{-1}=(g_{i,kl})_{k,l=1,2,\ldots,M_i}$.
Then we have that $E\{\mf_{kl}(M_i)\}=\b0$ for $k,l$
$=1,\ldots,M_i$, and
%If $\mf_{kl}(m_i)=\mf_{12}(m_i)$ for $1\leq k\neq l\leq m_i$, then we have that
$$E\{\meta_i(\bbeta^*)\}= E\{\mf_{11}(M_i) \sum_{k=1}^{M_i}g_{i,kk}+\sum_{1\leq k\neq l\leq M_i}g_{i,kl}\mf_{kl}(M_i) \}.$$

\end{Proposition}
When the cluster size is non-informative, $\mf_{kl}(M_i)$ are independent of the cluster size $M_i$. By the law of iterated expectations, we have $\mf_{kl}(M_i) \equiv\b0$ for $k,l=1,\ldots,M_i$ from equation (\ref{MT1}). This leads to $E\{\meta_i(\bbeta^*)\}=\b0$ for any predetermined working correlation matrix, making the GEE method is robust to the choice of  the working correlation matrix in such cases. However, when the cluster size is informative,  $\mf_{kl}(M_i)\neq\b0$ for $k,l=1,\ldots,M_i$ as $\mf_{kl}(M_i)$ depends $M_i$. From  Proposition \ref{prop1}, the solution set to $E\{\meta_i(\bbeta^*)\}=\b0$ is given by $$\mathcal{M}=\biggl\{\bG_{i}:\, E\{\mf_{11}(M_i) \sum_{k=1}^{M_i}g_{i,kk}\}+E\{\sum_{1\leq k\neq l\leq M_i}g_{i,kl}\mf_{kl}(M_i) \}=\b0\biggr\}.$$
Thus, the validness of GEE method depends on whether the predetermined working correlation matrix belongs to the set $\mathcal{M}$. The right choice of working correlation matrix depends on the unknown form of $\mf_{kl}(M_i)$, $k,l=1,\ldots,M_i$ and unknown distribution of $M_i$. There is no guarantee that the predetermined working correlation matrix can be included in  $\mathcal{M}$, and in some cases, $\mathcal{M}$ may even be an empty set. Therefore, the GEE method is not robust to the choice of  the working correlation matrix when ICS is present. This motivates us to develop a novel method, which is robust to  the choice of  the working correlation matrix
whether the cluster size is informative or not.

\subsection{Weighted Generalized Estimating Equations}\label{Sec2.2}
In this section, we propose a novel weighted estimating equation method to extend robustness of the  GEE framework  to scenarios where the cluster size is informative.
Let
$\bW_i=(\omega_{i,kl})_{k,l=1}^{M_i}$ be a $M_i\times M_i$ matrix. To eliminate the effect of $M_i$, we consider the following weighted estimating equations:
\begin{equation}\label{wgee}
\tilde{\meta}_i(\bbeta)=\bX_i^T\mF_i(\bbeta)\bigl\{\bG_i^{-1}\circ\bW\bigr\}\bPhi_i^{-1/2}(\bbeta)
(\bY_{i}-\bmu_i(\bbeta)),
\end{equation}
where $\circ$ denotes the element-wise product.
Our  idea is to find the optimal weighted matrix $\bW$ such that
$E\{\tilde{\meta}_i(\bbeta^*)\}= \b0$. Similar to the results in Proposition \ref{prop1}, we can easily obtain that
$$E\{\tilde{\meta}_i(\bbeta^*)\}= E\{\mf_{11}(M_i) \sum_{k=1}^{M_i}\omega_{i,kk}g_{i,kk}\}+E\{\sum_{1\leq k\neq l\leq M_i}\omega_{i,kl} g_{i,kl}\mf_{kl}(M_i)\}.$$
Since  it is difficult to capture the random relationship between $\sum_{1\leq k\neq l\leq M_i}\omega_{i,kl} g_{i,kl}\mf_{kl}(M_i)$ and $\mf_{11}(M_i) \sum_{k=1}^{M_i}\omega_{i,kk}g_{i,kk}$, we turn our focus to choosing $\bW_i$ to satisfy the following two equations:
\begin{equation}\label{TGEE}
E\{\mf_{11}(M_i) \sum_{k=1}^{M_i}\omega_{i,kk}g_{i,kk}\}=\b0 \ \ {\rm and} \ \
E\{\sum_{1\leq k\neq l\leq M_i}\omega_{i,kl} g_{i,kl}\mf_{kl}(M_i)\}=\b0.
\end{equation}
For the first equation, if we pick $\omega_{i,kk}= \{\sum_{k=1}^{M_i}g_{i,kk}\}^{-1}$ for $k=1,\ldots,M_i$, then $$E\{\mf_{11}(M_i) \sum_{k=1}^{M_i}\omega_{i,kk}g_{i,kk}\}
=E\{\mf_{11}(M_i)\}=\b0.$$
Let $\{B_{i,kl},k,l=1,\ldots,M_i\}$ be a random sample generated from  Rademacher distribution. The generation process of $B_{i,kl}$  is independent of $\{(Y_{ij},\bX_{ij}),j=1,\ldots,M_i\}$.
If we pick $\omega_{i,kl}=\{\sum_{1\leq k\neq l\leq m_i}g_{i,kk}\}^{-1}B_{i,kl} $ for $1\leq k\neq l\leq M_i$, since $E(B_{i,kl}\mid M_i)=0$, then
$$E\{\sum_{1\leq k\neq l\leq M_i}\omega_{i,kl} g_{i,kl}\mf_{kl}(M_i)\}
= E\{\sum_{1\leq k\neq l\leq M_i}\tilde{\mf}_{kl}(M_i) E(B_{i,kl}\mid M_i)\}=\b0,$$
where $\tilde{\mf}_{kl}(M_i)=\{\sum_{1\leq k\neq l\leq M_i} g_{i,kl}\}^{-1}g_{i,kl}
\mf_{kl}(M_i)$.
Then one solution to \eqref{TGEE} is
$$
\widetilde{\bW}=(\tilde\omega_{i,kl})_{k,l=1}^{M_i}, \ \
\tilde\omega_{i,kl}=\biggl\{\begin{array}{lc}
\{\sum_{k=1}^{M_i}g_{i,kk}\}^{-1}, & {\rm if}  \ \ k=l=1,\ldots,M_i,\\
\{\sum_{1\leq k\neq l\leq M_i} g_{i,kl}\}^{-1}B_{i,kl}, 	&  \ \ {\rm if} \ \  1\leq k\neq l\leq M_i. %{\sum_{1\leq k\neq l\leq m_i} g_{i,kl}\}^{-1}
	\biggr.
\end{array}$$
Substituting the weighted matrix $\widetilde{\bW}$ into \eqref{wgee} the leads to the following weighted estimating equations:
\begin{align}\label{MT4}
	\bQ_n\overset{\text{def}}{=}n^{-1}\sum_{i=1}^n\tilde{\meta}_i(\bbeta)
	=\b0,
\end{align}
where $\tilde{\meta}_i(\bbeta)=\bX_i^T\mF_i(\bbeta)\tilde\bG_i^{-1}\bPhi_i^{-1/2}(\bbeta)
(\bY_{i}-\bmu_i(\bbeta))$ with $\tilde\bG_i^{-1}=\bG_i^{-1}\circ\widetilde{\bW}$. The solution to equation (\ref{MT4}) provides an unbiased estimation of $\bbeta^*$, which we refer to as the WGEE estimator. It is worth noting that $\tilde\bG^{-1}_i$
is not the inverse of the working correlation matrix. However, the WGEE method can still be applied in the classical case where outcomes are independent of cluster size, $m_i$, although it may not be optimal as expected when cluster size is noninformative.

\subsection{Penalized Weighted Generalized Estimating Equations}\label{Sec2.3}
Regularized methods are fundamental for simultaneous estimation and variable selection in high dimensional regression. Given the advantageous properties of WGEE, we propose a PWGEE approach for analyzing high-dimensional longitudinal data when the cluster size is informative. The PWGEE objective function is given by:
\begin{align*}
	{\bQ}^P_n(\bbeta)&={\bQ}_n(\bbeta)-\brho_{\lambda_{n}}({|\bbeta|})\circ \sgn(\bbeta)\\
	&=n^{-1}\sum_{i=1}^n\tilde{\meta}_i(\bbeta) {-} \brho_{\lambda_{n}}({|\bbeta|})\circ \sgn(\bbeta),
\end{align*}
where $\brho_{\lambda_{n}}({|\bbeta|})=(\rho_{\lambda_{n}}(|\beta_1|),\ldots,\rho_{\lambda_{n}}(|\beta_{p_n}|))^T$, $\sgn(\bbeta)=(\sgn(\beta_1),\ldots,\sgn(\beta_{p_n}))^T$, with
$\sgn(t)=1(t>0)-1(t<0)$ and $\sgn(0)=0$. The symbol $\circ$ denotes the component-wise product. To achieve the consistency in variable selection, certain assumptions on the penalty function $\rho_{\lambda_{n}}(t)$ are required.
\begin{enumerate}
	\item [(A1)] Let $\bar{\rho}(t,\lambda_n)=\lambda_{n}^{-1}{\rho}_{\lambda_n}(t)$ for $\lambda_n>0$.
	Assume that (i) $\bar{\rho}(t,\lambda_n)$ is continuous with $\bar{\rho}(t,\lambda_n)\geq 0$ for $(0,\infty)$; (ii) $\bar{\rho}(t,\lambda_n)$ is increasing in $\lambda_{n}\in (0,\infty)$;
	(iii) $\bar{\rho}(0+,\lambda_n)$ is independent of $\lambda_{n}$.
\end{enumerate}
The class of penalty functions described above has been extensively studied in the literature \citep{FL11, Sel19}. The lasso penalty is a limiting case within this class. Both the SCAD penalty \citep{Fan2001} and the MCP penalty \citep{Zhang2010} satisfy the conditions outlined in (A1).

Given that $\brho_{\lambda_{n}}(|\bbeta|)\circ \sgn(\bbeta)$ is not continuous at $\b0$, the equation $\bQ^P_n(\bbeta)=\b0$ does not have an exact solution. Instead, an approximate solution $\hat\bbeta$ exists that satisfies the equation
$$\bQ_n(\hat\bbeta){-}\bv=\b0,$$
where $\bv=(v_1,\ldots,v_{p_n})$ with $v_j=\rho_{\lambda_{n}}(|\hat\beta_{j}|)\sgn(\hat\beta_{j})$ for $\hat\beta_j\neq 0$, and
$v_j\in [-\lambda_{n}\bar{\rho}(0+,\lambda_n),\lambda_{n}\bar{\rho}(0+,\lambda_n)]$ for
$\hat\beta_j=0$. This implies that
$\bQ_n(\hat\bbeta)=O(\lambda_{n})$.

For any subset {$\mathcal{S}\subset \{1,2,\ldots,p_n\}$}, let $(\bQ_n(\bbeta))_{\mathcal{S}}$ be the subvector formed by the components in $\mathcal{S}$. Define
$$\bD_{n}(\bbeta,\mathcal{S})=\diag(\rho_{\lambda_{n}}(|\beta_1|)/(c+|\beta_1|),\ldots,\rho_{\lambda_{n}}(|\beta_{\mathcal{S}}|)/(c+|\beta_{\mathcal{S}}|)),$$ where $c$ is a small constant, and
$$\bH_{n}(\bbeta,\mathcal{S})=n^{-1}\sum_{i=1}^n
\frac{\partial\bmu_i(\bbeta)}{\partial \bbeta_{\mathcal{S}}^T}
\bPhi_i^{-1/2}(\bbeta)\tilde\bG_i^{-1}\bPhi_i^{-1/2}(\bbeta)
\frac{\partial\bmu_i(\bbeta)}{\partial \bbeta_{\mathcal{S}}}.$$
We now introduce an algorithm to implement the PWGEE. The algorithm is based on the Newton-Raphson method, following the approach in \cite{Fan2001}. The steps of the proposed algorithm are outlined as follows.

\begin{algorithm}[h!]
	%\floatname{algorithm}{Algorithm}% žüžÄËã·šÇ°×ºÃû³Æ
	%\renewcommand{\algorithmicrequire}{\textbf{Input:}}% žüžÄÊäÈëÃû³Æ
	%\renewcommand{\algorithmicensure}{\textbf{Output:}}% žüžÄÊä³öÃû³Æ
	\footnotesize
	\caption{Quasi-newton raphson algorithm}
	\label{alg1}
	\begin{algorithmic}[1]
		\REQUIRE Data $\{(Y_{ij},\bX_{ij}), j=1,\ldots,M_i\}_{i=1}^n$, initial value $\hat\bbeta^{(0)}$, tuning parameter $\lambda_{n}$, the number of
		iterations $L$.
		\STATE Set $\mathcal{S}=\{1,2,\ldots,p_n\}$.
		
		\STATE For $k=2: L$ do
		\begin{itemize}
			\item []  2.1 Compute $\bH_{n}(\hat\bbeta^{(k-1)},\mathcal{S})$,
			$\bD_{n}(\hat\bbeta^{(k-1)},\mathcal{S})$ and $\bQ_n(\hat\bbeta^{(k-1)})$.
			\item [ ] 2.2 Update
			$\hat\bbeta_{\mathcal{S}}^{(k)}=\hat\bbeta_{\mathcal{S}}^{(k-1)}+\{\bH_{n}(\hat\bbeta^{(k-1)},\mathcal{S})
			+\bD_{n}(\hat\bbeta^{(k-1)},\mathcal{S})\}^{-1}
			\times(\bQ_n(\hat\bbeta^{(k-1)})-\rho_{\lambda_{n}}(|\hat\bbeta^{(k-1)}|)\sgn(\hat\bbeta^{(k-1)}))_{\mathcal{S}}$
			and
			$\hat\bbeta_{\mathcal{S}^c}^{(k)}=\b0$.
			\item [] 2.3 Update $\mathcal{S}=
			\{j: n^{-1}\sum_{i=1}^n |\tilde{\eta}_{ij}(\hat\bbeta^{(k)})|>
			\lambda_{n}\bar{\rho}(0+,\lambda_n)\}$.
			\item [] 2.4 If $\sum_{d=1}^{p_n}|\hat\bbeta^{(k)}_j-\hat\bbeta^{(k-1)}_j|\leq 10^{-15}$, break.
		\end{itemize}
		\STATE Output $\hat\bbeta=\hat\bbeta^{(k)}$.

	\end{algorithmic}
\end{algorithm}

To reduce the computational burden,  we introduce a set $\mathcal{S}$ in step 2.3. If the terminal condition $ n^{-1}\sum_{i=1}^n |\tilde{\eta}_{ij}(\hat\bbeta^{(k)})|\leq\lambda_{n}\bar{\rho}(0+,\lambda_n)$ is met, we can cease updating $\hat\beta_{j}^{(k)}$ and set $\hat\beta_{j}=0$ for $j\in \mathcal{S}^c$. This approach accelerates the update process in step 2.2. In practice, a simplified threshold level can be used for convergence, such as $\mathcal{S}=\{j: |\hat\beta_{j}^{(k)}|>10^{-3}\}$.

\section{Asymptotic Results} \label{Sec3}
\hspace{0.5cm} In this section, we establish the asymptotic properties of the PWGEE estimator
$\hat\bbeta$ within a generalized framework which  accommodates ICS.

For simplicity, we first introduce some notations. Let $\bbeta^*=(\{\bbeta_{1}^*\}^T,\{\bbeta_{2}^*\}^T)^T$ be  the true parameter vector in (\ref{MT1}), where $\bbeta_{2}^*=\b0$. Define $\text{supp}(\bbeta^*)$ as the set of non-zero indices of $\bbeta^*$. The covariate matrix $\bX_i$ is partitioned into two submatrices by columns, denoted as  $\bX_i=(\overline{\bX}_{i1},\overline{\bX}_{i2})$. Here, $\overline{\bX}_{i1}$ represents the submatrix of $\bX_i$ corresponding to $\bbeta_{1}^*$, and $\overline{\bX}_{i2}$ represents the submatrix of $\bX_i$ corresponding to $\bbeta_{2}^*$. Similarly, the WGEE function $\tilde{\meta}_{i}(\bbeta)$ is  partitioned as
$\tilde{\meta}_{i}(\bbeta)=(\{\tilde{\meta}_{i1}(\bbeta)\}^T,\{\tilde{\meta}_{i2}(\bbeta)\}^T)^T$. Denote by
$\nabla \tilde{\meta}_{il}(\bbeta) = \partial\tilde{\meta}_{il}(\bbeta) /\partial\bbeta_{1}^T $ the derivative of $\tilde{\meta}_{il}(\bbeta)$ with respect to $\bbeta_{1}^T $ for $l=1,2$. Let $\be_j \in \mathbb{R}^{p_n}$ be the unit vector with the $j$-th component being $1$. For any $\mathcal{S}\subset\{1,2,\ldots,p_n\}$
, $\overline{\bX}_{i\mathcal{S}}$ represents the submatrix of $\bX_i$ formed by the column indices in $\mathcal{S}$. Furthermore, we denote the dimension of $\bbeta_{1}^*$  as $s_n$.

To establish both variable selection consistency and estimation consistency, the following regularity conditions are required:
\begin{enumerate}
	\item[(A2)] (i) There exists positive constants $c_1$ and $c_2$ such that
	$$c_1< \lambda_{\min}\bigl(-E\bigl(\nabla \tilde{\meta}_{i1}(\bbeta^*)\bigr)\bigr)$$ and  $$\lambda_{\max}\bigl(E\bigl(\tilde{\meta}_{i1}(\bbeta^*)\{\tilde{\meta}_{i1}(\bbeta^*)\}^T\bigr)\bigr)<c_2;$$
	(ii) There exists some positive constant $\delta>0$ such that
	$$E|b(\bX_{il}^T\bbeta^*) c'(\bX_{it}^T\bbeta^*)|^{2+\delta}=O(1)$$ and
	$$E | b'(\bX_{il}^T\bbeta^*) c(\bX_{it}^T\bbeta^*)|^{2+\delta}=O(1),$$  uniformly for $i=1,\ldots,n$, $l,t=1,\ldots,M_i$, where $b(t)=\mu(t)\phi^{-1/2}(t)$ and $c(t)=\phi^{-1/2}(t)$.
	\item [(A3)] The following results hold with probability going to one:
	\begin{equation}\label{MT7}
		\max_{j=s_n+1,\ldots,p_n}\lambda_{\max}\bigl\{n^{-1}\sum_{i=1}^n \overline{\bX}_{i\mathcal{S}_j}^T\mF_i(\bbeta^*)
		\tilde\bG_i^{-1}\mF_i(\bbeta^*)\overline{\bX}_{i\mathcal{S}_j}\bigr\}
		=O(n)
	\end{equation}
	and
	\begin{align}\label{MT8}
		\max_{j=1,\ldots,p_n}\max_{\bldsr{\bbeta} \in\mathbb{C}}\max_{\|\ba\|_2=1}&n^{-1}\sum_{i=1}^{n}|\ba^T\big\{\partial \bigl(\be_j^T\tilde\meta_{i}(\bbeta)\bigr)/\partial \bbeta_{1}\partial \bbeta_{1}^T
		\bigr\}\ba| \nonumber\\
		&=O(\sqrt{\log(n)\log(np_n)})
	\end{align}
	where $\mathcal{S}_j=\text{supp}(\bbeta^*)\cup\{j\}$, and
	$\mathbb{C}=\{\bbeta: \|\bbeta_{1}-\bbeta_{1}^*\|_2\leq \tau \sqrt{s_n/n}, \bbeta_{2}=\b0\}$ for some positive constant $\tau$.
	
	\item [(A4)] (i) There exists constants $c_3>0$ and $c_4>0$ such that
	$E\{\exp(c_3|Y_{ij}-\mu(\bX_{ij}^T\bbeta^*)|)\mid \bX_{ij},M_i\}\leq c_4$ and $E\{\exp(c_3|X_{ij,k}|)\}\leq c_4$, uniformly in $i=1,\ldots,n$,
	$j=1,\ldots,M_i$, $k=1,\ldots,p_n$;
	(ii)  The maximum cluster size $M=\max\limits_{i=1,\cdots, n} M_i$ is bounded away from $\infty$;
	(iii) $\bG^{-1}$ is a positive definite matrix with eigenvalues bounded away
	from $0$ and infinity;

	\item [(A5)] Let $d_n=\min\limits_{j=1,\ldots,s_n}|\beta_j^*|/2$. Assume:
	(i) $\max \{s_n^{1/2}n^{-1/2}, \log(p_n)n^{-1/2}\}\ll \lambda_{n} \ll d_n $ and $\rho_{\lambda_n}(d_n)$
	$=O(n^{-1/2})$ ;
	(ii) $s_n=o(n^{1/3})$, $s_n\log^3(p_n)=o(n)$ and $s_n^2\log(n)\log(np_n)=o(n)$.

\end{enumerate}

\noindent \textbf{ Discuss on Conditions}.
Condition (A2)(i) requires that the Fisher information matrix and covariance matrix of weighted estimating equations are positive definite, with bounded eigenvalues. This is a common assumption for independent data \citep{FP04}.
Condition (A2)(i) is implied by
\begin{equation}\label{key1}
	\max_{j=s_n+1,\ldots,p_n}\lambda_{\max}\bigl\{ n^{-1}\sum_{i=1}^n \overline{\bX}_{i\mathcal{S}_j}^T\overline{\bX}_{i\mathcal{S}_j}\bigr\}=O(1),
\end{equation}
and
\begin{align}\label{key2}
	\max_{ \bldsr{\bbeta}\in\mathbb{C}}\max_{i=1,\ldots,n}\max_{j=1,\ldots,M_i}\mu_{k}(\bX_{ij}^T\bbeta)&=O(1),\nonumber\\
	\max_{\bldsr{\bbeta}\in\mathbb{C}}\max_{i=1,\ldots,n}\max_{j=1,\ldots,M_i}\{\phi_{k-1}(\bX_{ij}^T\bbeta)\}^{-1}&=O(1),
\end{align}
where $\mu_{k}$ denotes the $k$-th derivative of $\mu$, and  $\phi_{k-1}$ is the $(k-1)$-th derivative of $\phi$, with $k=1,2,3$. Here, $\phi_0 \equiv\phi$ and $\mu_{1}=\mu'$.
Condition (A3) is the technique condition for establishing the sparsity of the PWGEE estimator. For independent data, Condition (A3)  is implied by Condition 4 in \cite{FL11}. For clustered data, \eqref{MT7} is directly implied by assumptions \eqref{key1} and \eqref{key2}, and  \eqref{MT8} also holds under assumptions \eqref{key1} and \eqref{key2} (Lemma 5 in the Supplementary). Assumptions \eqref{key1} and \eqref{key2} follow from \cite{Wang2012} and
\cite{Wang2011} in the context of longitudinal analysis.
Under assumption \eqref{key2}, Condition (A2)(i) is easily satisfied.

Condition (A4)(i) assumes that covariates are sub-Gaussian, which is a weaker condition than the bounded assumption (A1) in \cite{Wang2012}.
The assumption for the error term $Y_{ij}-\mu(\bX_{ij}^T\bbeta^*)$  in (A4)(i) is satisfied for Gaussian, sub-Gaussian, Binomial and Poisson distributions, which is commonly used in the literature \citep{Wang2012,Wang2011, FL11}. Conditions (A4)(ii) and (A4)(iii) are standard assumptions for the GEE method in longitudinal data analysis \citep{Wang2011,Wang2012}.

Condition (A5)(i) imposes some restrictions on the tuning parameter $\lambda_{n}$.  Specifically, when $d_n\gg \lambda_{n}$, we have $\rho_{\lambda_{n}}(d_n)
=0$ for SCAD and MCP penalties. To establish variable selection consistency, \cite{Wang2012} assumed that $d_n\gg \lambda_{n}$,
$\lambda_{n}\gg s_nn^{-1/2}$, and $\lambda_{n}\gg {p_n}^{1/4}n^{-1/2}$, which are stronger requirements than those in (A5)(i).
From Condition (A5)(ii), we allow $\log(p_n)=O(n^{2/9})$, meaning $p_n$ can diverge exponentially with the sample size $n$.
By contrast, \cite{Wang2012} restricted $p_n=O(n)$.

\begin{theorem}\label{Th1}(Consistency)
	Assume Conditions ($A_1$)--($A_5$) hold. There exists an approximate solution to the PWGEE, denoted as $\hat\bbeta=(\hat\bbeta_{1}^T,\hat\bbeta_{2}^T)^T$ such that
	with probability tending to one, $\hat\bbeta_{2}=\b0$ and
	$\|\hat\bbeta_{1}-\bbeta_{1}^*\|_{2}=O_{p}(\sqrt{s_n/n})$.
	
\end{theorem}
Theorem
%<*Th1>
\ref{Th1}
%</Th1>
establishes the consistency of the PWGEE estimator in the sense that $\hat\bbeta_{2}=\b0$ with probability going to one and $\|\hat\bbeta_{1}-\bbeta_{1}^*\|_{2}=O_{p}(\sqrt{s_n/n})$. This result is often referred to as the  weak oracle property in penalized regression \citep{FL11}.

In the following, we establish the asymptotic distribution of the PWGEE estimator. Before that, we introduce the following technical condition.
\begin{enumerate}
	\item[(A6)] $\rho_{\lambda_{n}}(d_n)=o(s_n^{-1/2}n^{-1/2})$ and
	$s_n^{1/4}\sum_{i=1}^nE\|\bC^{-1}\mathcal{X}_i\|_{2}^3=o(1)$, where $\mathcal{X}_{i}=n^{-1/2}\bigl\{E\nabla\tilde\meta_{i1}(\bbeta^*)\bigr\}^{-1}$
	$\tilde{\meta}_{i1}(\bbeta^*)$ and $\bC^2=\sum_{i=1}^n\text{Cov}(\mathcal{X}_{i})$.
\end{enumerate}

When $d_n=o(\lambda_{n})$, $\rho_{\lambda_{n}}(d_n)=0$ for SCAD and MCP penalties, which implies that $\rho_{\lambda_{n}}(d_n)=o(s_n^{-1/2}n^{-1/2})$ is easily satisfied. This assumption $ns_n^{1/4}E\|\bC^{-1}\mathcal{X}_i\|_{2}^3=o(1)$  serves as Lyapunov condition for  the Central Limit Theorem (CLT).
Under Conditions (A1) and (A4), along with assumptions \eqref{key1} and \eqref{key2}, we have
\begin{align*}
	&s_n^{1/4}\sum_{i=1}^nE\|\bC^{-1}\mathcal{X}_i\|_{2}^3\\
	&\leq O(1)s_n^{1/4} n^{-3/2}\sum_{i=1}^n \bigl\{\sum_{j=1}^{M_i} (Y_{ij}-\mu(\bX_{ij}^T\bbeta^*))\bigr\}^{3/2}\\
	&=O(s_n^{1/4}n^{-1/2}).
\end{align*}
This ensures that $s_n^{1/4}\sum_{i=1}^nE\|\bC^{-1}\mathcal{X}_i\|_{2}^3=o(1)$.

\begin{theorem}(Asymptotic normality)\label{Th2}
	Assume that Conditions (A1)--(A6) hold. With probability going to one,
	there exists an approximate solution to the PWGEE: $\hat\bbeta=(\hat\bbeta_{1}^T,\hat\bbeta_{2}^T)^T$ such that
	$\hat\bbeta_{2}=\b0$  and
	$$\sqrt{n}\bA_n\bC^{-1}(\hat\bbeta_{1}-\bbeta_{1}^*) \overset{D}\to N(\b0,\bA)$$
	where  $\bC^2=
	\bigl\{E\nabla\tilde\meta_{i1}(\bbeta^*)\bigr\}^{-1}E\bigl\{\tilde{\meta}_{i1}(\bbeta^*)\tilde{\meta}_{i1}(\bbeta^*)^T\bigr\}\bigl\{E\nabla\tilde\meta_{i1}(\bbeta^*)\bigr\}^{-1}$, and
	$\bA_n$ is a $q\times s_n$ matrix such that $\bA_n\bA_n^T \to \bA$, with $\bA$  being a symmetric positive definite matrix.
	
\end{theorem}
By Theorem \ref{Th2}, the PWGEE estimator  exhibits  the strong oracle property, meaning it correctly identifies the true zero coefficients as zero and estimates the nonzero coefficients with the same efficiency as the oracle estimator, assuming prior knowledge of the true model.

\begin{Remark}{\rm
		By Theorem \ref{Th2}, variable selection consistency and asymptotic normality hold, regardless of whether the cluster size is informative. In the framework of our PWGEE, we  allow the cluster size to be informative,  meaning that $E(Y_{ij}\mid \bX_{ij},M_i)\neq E(Y_{ij}\mid \bX_{ij})$. This extends the traditional GEE framework to accommodate more complex scenarios, ensuring that our penalized WGEE method performs robustly even when the cluster size is informative.
	}
\end{Remark}

\begin{Remark}{\rm
		When the cluster size is non-informative, \cite{Wang2012} demonstrated that the strong oracle property held for PGEE under the assumption $p_n=o(n^{\alpha})$, $0<\alpha< 4/3$.
		Under our framework, with some mild assumptions, we allow the dimension $p_n$ to grow exponentially with the sample size $n$, such that
		$p_n=\exp(n^{\alpha})$, $0<\alpha\leq 2/9$.
	}
\end{Remark}

\section{Simulations}\label{Sec4}

In this section, we execute several numerical experiments to evaluate the performance of our PWGEE method in model selection and estimation for high-dimensional longitudinal data. The experiments include analyses of correlated continuous and Poisson responses.

We compare the performance of our proposed PWGEE and its oracle version (i.e. the WGEE with the true marginal regression model was known in advance) with several alternatives: the naive Lasso, which treats all responses as independent and fits a generalized linear model using penalized maximum likelihood; the PGEE method proposed by \cite{Wang2012}; and the oracle GEE (i.e., the GEE with the true marginal regression model known in advance). We use the R package \texttt{PGEE} for the penalized GEE and  oracle GEE procedure. Three commonly used working correlation structures are considered:  independence, exchangeable (equally correlated) and AR(1) (autocorrelation). To facilitate the comparison of various methods under different correlation scenarios,   we use the suffixes ".indep," ".exch" and ".ar1" to represent the three aforementioned correlation structures, respectively.
Following \cite{Wang2012}, we  use fourfold cross-validation to select the tuning parameter for the SCAD penalty function in the PWGEE and PGEE, and classify a coefficient as zero if the magnitude of its estimated value is smaller than the threshold value $10^{-3}$.

We generate 100 longitudinal data sets for each setup.
To evaluate the screening performance, we calculate the following criteria: (1) True Positive (TP): the average number of true variables correctly identified; (2) False Positive (FP): the average number of unimportant variables mistakenly selected; (3) Coverage Rate (CR): whether the true variables are contained in the selected model, with a value of 1 indicating  $\mathcal{S}\subseteq\hat{\mathcal{S}}$, and 0 indicating otherwise; and (4) Mean Square Error (MSE): calculated as $\sum_{k=1}^{100}\Vert\hat{\bm{\beta}}^k-\bm{\beta}\Vert^2/100$, where $\hat{\bm{\beta}}^k$ is the estimate from the $k$th generated dataset.

The random cluster size $M_i$ takes value in the set $\{2,4,15\}$, with the following probability distribution: $P(M_i=2)=9/16$, $P(M_i=4)=3/8$, and $P(M_i=15)=1/16$.

In Example \ref{ex1} and Example \ref{ex2}, ICS is considered, that is, $E(Y_{ij}|\bX_{ij},M_i)\not=E(Y_{ij}|\bX_{ij})$.

\begin{Example}\label{ex1} We consider a linear regression model with ICS:
	\begin{equation*}
		Y_{ij}=\bX^T_{ij}\bm{\beta}+ U_{ij}+\epsilon_{ij},
	\end{equation*}
	for correlated normal responses, where $i=1,\dots,200$, $j=1,\dots,M_i$, the covariate vector $\bX_{ij}=(x_{ij,1},\dots,x_{ij,p_n})^T$ is $p_n$-dimensional, and the coefficient vector $\bm{\beta}=(2,-1,1,-1.5,0,\dots,0)^T$ represents the homogeneous effect. We define $U_{ij}=-1.5\bX^T_{ij}\bm{\beta}\cdot(1(M_i>4)-1/{4}^2)$. The covariates $(x_{ij,1},\dots,x_{ij,p_n})^T$ are generated from a multivariate normal distribution with mean $\b0$, marginal variance 1, pairwise correlation  of 0.5. The random errors  $(\epsilon_{i1},\dots,\epsilon_{iM_i})^T$ obey a multivariate normal distribution with marginal mean $\b0$, marginal variance 1, and an exchangeable correlation structure with parameter $\rho=0.5$.
	In this case, the cluster size is informative and the marginal mean of the response $\mu_{ij}=E(Y_{ij}|\bX_{ij})=\bX^T_{ij}\bm{\beta}$.
\end{Example}

\begin{table}[t]
		\renewcommand{\arraystretch}{0.8}
	\centering
	\caption{Model selection results for linear regression with ICS}
	\label{linear,nv=1}
	\vspace{0.2cm}
	\begin{tabular}{lccccc}
		\hline
		\multicolumn{1}{c}{$p_n$}	&Approaches & \multicolumn{1}{c}{TP} &\multicolumn{1}{c}{FP} &\multicolumn{1}{c}{CR}& \multicolumn{1}{c}{MSE}  \\
		\hline
		500&naive lasso&4.00(0.00)&1.83(1.85)&1.00(0.00)&1.048(0.290)\\		
		~&PGEE.indep&4.00(0.00) &3.79(1.55)  & 1.00(0.00)  &0.329(0.138)   \\
		~&PGEE.exch&4.00(0.00)&3.85(1.57) &1.00(0.00)  &0.391(0.153) \\
		~&PGEE.ar1&4.00(0.00)&3.87(1.56) &1.00(0.00)  &0.337(0.142)  \\		
		~&PWGEE.indep&4.00(0.00) &0.00(0.00)  & 1.00(0.00)  &0.016(0.012)   \\
		~&PWGEE.exch&4.00(0.00)&0.04(0.32) &1.00(0.00)  &0.033(0.028)  \\
		~&PWGEE.ar1&4.00(0.00)&0.02(0.14) &1.00(0.00)  &0.031(0.024)  \\		
		~&Oracle.GEE.indep&\multicolumn{1}{c}{-}&\multicolumn{1}{c}{-}&\multicolumn{1}{c}{-} &0.277(0.123)  \\
		~&Oracle.GEE.exch&\multicolumn{1}{c}{-}&\multicolumn{1}{c}{-} &\multicolumn{1}{c}{-} &0.343(0.140)  \\
		~&Oracle.GEE.ar1&\multicolumn{1}{c}{-}&\multicolumn{1}{c}{-} &\multicolumn{1}{c}{-}  &0.288(0.127)  \\		
		~&Oracle.WGEE.indep&\multicolumn{1}{c}{-}&\multicolumn{1}{c}{-} &\multicolumn{1}{c}{-} &0.016(0.012) \\
		~&Oracle.WGEE.exch&\multicolumn{1}{c}{-}&\multicolumn{1}{c}{-} &\multicolumn{1}{c}{-}  &0.032(0.026)  \\
		~&Oracle.WGEE.ar1&\multicolumn{1}{c}{-}&\multicolumn{1}{c}{-} &\multicolumn{1}{c}{-}  &0.030(0.023) \\		
		\hline
		1000&naive lasso&4.00(0.00)&2.05(2.35)&1.00(0.00)&1.134(0.281)\\
		~&PGEE.indep&4.00(0.00) &3.63(1.66)  & 1.00(0.00)  &0.346(0.123)   \\
		~&PGEE.exch&4.00(0.00)&3.67(1.66) &1.00(0.00)  &0.414(0.140) \\
		~&PGEE.ar1&4.00(0.00)&3.71(1.62) &1.00(0.00)  &0.356(0.125)  \\		
		~&PWGEE.indep&4.00(0.00) &0.01(0.10)  & 1.00(0.00)  &0.016(0.011)   \\
		~&PWGEE.exch&4.00(0.00)&0.02(0.14) &1.00(0.00)  &0.036(0.031) \\
		~&PWGEE.ar1&4.00(0.00)&0.02(0.14) &1.00(0.00)  &0.034(0.029)  \\		
		~&Oracle.GEE.indep&\multicolumn{1}{c}{-}&\multicolumn{1}{c}{-} &\multicolumn{1}{c}{-}  &0.284(0.106)  \\
		~&Oracle.GEE.exch&\multicolumn{1}{c}{-}&\multicolumn{1}{c}{-} &\multicolumn{1}{c}{-}  &0.351(0.126) \\
		~&Oracle.GEE.ar1&\multicolumn{1}{c}{-}&\multicolumn{1}{c}{-} &\multicolumn{1}{c}{-}  &0.297(0.109) \\		
		~&Oracle.WGEE.indep&\multicolumn{1}{c}{-}&\multicolumn{1}{c}{-} &\multicolumn{1}{c}{-} &0.015(0.010)  \\
		~&Oracle.WGEE.exch&\multicolumn{1}{c}{-}&\multicolumn{1}{c}{-} &\multicolumn{1}{c}{-}  &0.031(0.021) \\
		~&Oracle.WGEE.ar1&\multicolumn{1}{c}{-}&\multicolumn{1}{c}{-} &\multicolumn{1}{c}{-}  &0.030(0.020) \\	
		\hline
	\end{tabular}
\end{table}

Table~\ref{linear,nv=1} summarizes the model selection results and estimation accuracy for Example \ref{ex1}. All methods  yield sparse models, but the proposed PWGEE outperforms the other competing methods in terms of MSE. When ICS is present, the PWGEE shows significant improvements in estimation accuracy and selects almost zero redundant predictors compared with the PGEE under three working correlation structures. Our weighting operation on the working correlation matrix leads to at least tenfold reduction in MSE, indicating that PWGEE effectively mitigates the effects caused by ICS.   The PWGEE also achieves MSEs comparable to the oracle WGEE, particularly when observations within clusters are assumed to be independent. Among all prespecified working correlation structures, the proposed PWGEE achieves a zero FP at $p_n=500$ and the lowest MSE when repeated measurements are considered independent in the presence of ICS.
When the number of covariates increases to $p_n=1,000$, the estimation error of all methods rises in varying degrees. However, the PWGEE continues to demonstrate the best performance and robustness against ICS.

\begin{Example}\label{ex2}We consider the correlated Poisson responses $Y_{ij}$ with marginal mean $\mu_{ij}$, where ICS follows the model
	\begin{equation*}
		\log(\mu_{ij})=\bX^T_{ij}\bm{\beta}+ U_{ij}
	\end{equation*}
	for $i=1,\dots,200$, $j=1,\dots,M_i$. Here $\bX_{ij}=(x_{ij,1},\dots,x_{ij,p_n})^T$ is a $p_n$-dimensional vector of covariates,  $\bm{\beta}=(1,-0.8,0.9,-1,0,\dots,0)^T$ and $U_{ij}=\log (1+1.5|\bX^T_{ij}\bm{\beta}|\cdot (1(M_i>4)-1/{4}^2))$. Then, the marginal mean of the Poisson responses is $\mu_{ij}= E	(Y_{ij}|\bX_{ij})=\exp(\bX^T_{ij}\bm{\beta})$. The covariates $(x_{ij,1},\dots,x_{ij,p_n})^T$ are jointly normal distributed with mean $\b0$ and a compound symmetry covariance matrix with marginal variance 1 and correlation coefficient $0.5$. The correlated Poisson responses are generated by R package \texttt{wgeesel} with an exchangeable correlation matrix and parameter $\rho=0.5$.
\end{Example}

\begin{table}[t]
		\renewcommand{\arraystretch}{0.8}
	\centering
	\caption{Model selection results for Poisson regression with ICS}
	\label{poisson,nv=1}	
	\vspace{0.2cm}
	\begin{tabular}{lccccc}
		\hline
		\multicolumn{1}{c}{$p_n$}	&Approaches & \multicolumn{1}{c}{TP} &\multicolumn{1}{c}{FP} &\multicolumn{1}{c}{CR}& \multicolumn{1}{c}{MSE}  \\
		\hline
		500&naive lasso&4.00(0.00)&30.29(12.91)&1.00(0.00)&0.459(0.117)\\	
		~&PGEE.indep&4.00(0.00) &4.79(1.56)  & 1.00(0.00)  &0.172(0.106)   \\
		~&PGEE.exch&4.00(0.00)&5.58(0.67) &1.00(0.00)  &0.185(0.095)  \\
		~&PGEE.ar1&4.00(0.00)&5.02(1.21) &1.00(0.00) &0.157(0.099) \\		
		~&PWGEE.indep&4.00(0.00) &0.04(0.32)  & 1.00(0.00)  &0.036(0.038)  \\
		~&PWGEE.exch&4.00(0.00)&0.04(0.32) &1.00(0.00)  &0.045(0.044)\\
		~&PWGEE.ar1&4.00(0.00)&0.04(0.32) &1.00(0.00) &0.044(0.043) \\
		~&Oracle.GEE.indep&\multicolumn{1}{c}{-}&\multicolumn{1}{c}{-} &\multicolumn{1}{c}{-}   &0.174(0.550) \\
		~&Oracle.GEE.exch&\multicolumn{1}{c}{-}&\multicolumn{1}{c}{-} &\multicolumn{1}{c}{-}   &0.188(0.537) \\
		~&Oracle.GEE.ar1&\multicolumn{1}{c}{-}&\multicolumn{1}{c}{-} &\multicolumn{1}{c}{-}  &0.167(0.549) \\				
		~&Oracle.WGEE.indep&\multicolumn{1}{c}{-}&\multicolumn{1}{c}{-} &\multicolumn{1}{c}{-}   &0.037(0.044) \\
		~&Oracle.WGEE.exch&\multicolumn{1}{c}{-}&\multicolumn{1}{c}{-} &\multicolumn{1}{c}{-}   &0.045(0.050) \\
		~&Oracle.WGEE.ar1&\multicolumn{1}{c}{-}&\multicolumn{1}{c}{-} &\multicolumn{1}{c}{-}   &0.044(0.049) \\
		\hline 		
		1000&naive lasso&4.00(0.00)&34.78(12.18)&1.00(0.00)&0.514(0.122)\\		
		~&PGEE.indep&4.00(0.00) &4.57(1.45)  & 1.00(0.00)  &0.190(0.118)  \\
		~&PGEE.exch&4.00(0.00)&5.47(0.90) &1.00(0.00) &0.212(0.154) \\
		~&PGEE.ar1&4.00(0.00)&4.82(1.28) &1.00(0.00)  &0.166(0.110) \\
		~&PWGEE.indep&4.00(0.00) &0.01(0.10)  & 1.00(0.00)  &0.041(0.051) \\
		~&PWGEE.exch&4.00(0.00)&0.03(0.22) &1.00(0.00) &0.051(0.055) \\
		~&PWGEE.ar1&4.00(0.00)&0.03(0.22) &1.00(0.00)  &0.050(0.055) \\
		~&Oracle.GEE.indep&\multicolumn{1}{c}{-}&\multicolumn{1}{c}{-} &\multicolumn{1}{c}{-}  &0.124(0.091) \\
		~&Oracle.GEE.exch&\multicolumn{1}{c}{-}&\multicolumn{1}{c}{-} &\multicolumn{1}{c}{-}   &0.145(0.100) \\
		~&Oracle.GEE.ar1&\multicolumn{1}{c}{-}&\multicolumn{1}{c}{-} &\multicolumn{1}{c}{-}   &0.118(0.089) \\				
		~&Oracle.WGEE.indep&\multicolumn{1}{c}{-}&\multicolumn{1}{c}{-} &\multicolumn{1}{c}{-}   &0.034(0.041) \\
		~&Oracle.WGEE.exch&\multicolumn{1}{c}{-}&\multicolumn{1}{c}{-} &\multicolumn{1}{c}{-}   &0.043(0.049) \\
		~&Oracle.WGEE.ar1&\multicolumn{1}{c}{-}&\multicolumn{1}{c}{-} &\multicolumn{1}{c}{-}   &0.042(0.047) \\				
		\hline
	\end{tabular}
\end{table}

Correlated count responses convey less information than continuous data, complicating the identification of significant covariates and the estimation of  parameter coefficients. The results from Example \ref{ex2}, shown in Table~\ref{poisson,nv=1}, demonstrate that our PWGEE performs optimally when handling high-dimensional correlated Poisson data, especially in the presence of ICS. Although PWGEE does not achieve the same level of estimation accuracy as their oracle counterparts, given the nature of the responses, they still significantly outperform PGEE by minimizing the impact of ICS, yielding lower MSEs and fewer FPs. Much like in linear regression, PWGEE exhibits optimal performance when observations are treated as independent—a pattern also observed in its oracle version. By contrast, the naive versions of Lasso identify numerous redundant covariates, leading to inconsistent model selection and unreliable estimation results.

In Example \ref{ex3} and Example \ref{ex4}, we consider scenarios without ICS.

\begin{Example}\label{ex3} We use a linear regression model for correlated normal responses, defined as
	\begin{equation*}
		Y_{ij}=\bX^T_{ij}\bm{\beta}+\epsilon_{ij},
	\end{equation*}
	where $i=1,\dots,200$, $j=1,\dots,M_i$, $\bX_{ij}=(x_{ij,1},\dots,x_{ij,p_n})^T$ is a $p_n$-dimensional covariate vector, and $\bm{\beta}=(2,-1,1,-1.5,0,\dots,0)^T$ represents the homogeneous coefficient effects. The covariates are generated as shown in Example \ref{ex1}. The random error vector $(\epsilon_{i1},\dots,\epsilon_{iM_i})^T$ follows a multivariate normal distribution with  mean $\b0$, marginal variance 1 and an exchangeable correlation structure with parameter $\rho=0.5$.
\end{Example}

\begin{Example}\label{ex4}
	We model the correlated Poisson responses with marginal mean $\mu_{ij}$ satisfying
	\begin{equation*}
		\log(\mu_{ij})=\bX^T_{ij}\bm{\beta},
	\end{equation*}
	where $i=1,\dots,200$, $j=1,\dots,M_i$, $\bX_{ij}=(x_{ij,1},\dots,x_{ij,p_n})^T$ is a $p_n$-dimensional covariate vector, and $\bm{\beta}=(1,-0.8,0.9,-1,0,\dots,0)^T$.  The covariates are generated as those in Example 2. The within-cluster correlation is set as $\rho=0.5$, assuming an exchangeable correlation structure.
\end{Example}

\begin{table}[t]
	\renewcommand{\arraystretch}{0.8}
	\centering
	\caption{Model selection results for Linear regression without ICS}
	\label{linear,nv=0}
	\vspace{0.2cm}
	\begin{tabular}{lccccc}
		\hline
		\multicolumn{1}{c}{$p_n$}	&Approaches & \multicolumn{1}{c}{TP} &\multicolumn{1}{c}{FP} &\multicolumn{1}{c}{CR}& \multicolumn{1}{c}{MSE}  \\
		\hline 		
		500&naive lasso&4.00(0.00)&2.15(1.79)&1.00(0.00)&0.125(0.052)\\		
		~&PGEE.indep&4.00(0.00) &2.67(2.45)  & 1.00(0.00)  &0.030(0.028) \\
		~&PGEE.exch&4.00(0.00)&2.99(2.26) &1.00(0.00)  &0.017(0.015)  \\
		~&PGEE.ar1&4.00(0.00)&2.82(2.19) &1.00(0.00)  &0.019(0.017) \\		
		~&PWGEE.indep&4.00(0.00) &0.00(0.00)  & 1.00(0.00)  &0.012(0.009) \\
		~&PWGEE.exch&4.00(0.00)&0.11(0.45) &1.00(0.00)  &0.027(0.021)\\
		~&PWGEE.ar1&4.00(0.00)&0.09(0.40) &1.00(0.00)  &0.027(0.022) \\		
		~&Oracle.GEE.indep&\multicolumn{1}{c}{-}&\multicolumn{1}{c}{-} &\multicolumn{1}{c}{-}   &0.010(0.008) \\
		~&Oracle.GEE.exch&\multicolumn{1}{c}{-}&\multicolumn{1}{c}{-} &\multicolumn{1}{c}{-}   &0.006(0.005) \\
		~&Oracle.GEE.ar1&\multicolumn{1}{c}{-}&\multicolumn{1}{c}{-} &\multicolumn{1}{c}{-}   &0.007(0.006)  \\		
		~&Oracle.WGEE.indep&\multicolumn{1}{c}{-}&\multicolumn{1}{c}{-} &\multicolumn{1}{c}{-}   &0.012(0.009)  \\
		~&Oracle.WGEE.exch&\multicolumn{1}{c}{-}&\multicolumn{1}{c}{-} &\multicolumn{1}{c}{-}   &0.026(0.020) \\
		~&Oracle.WGEE.ar1&\multicolumn{1}{c}{-}&\multicolumn{1}{c}{-} &\multicolumn{1}{c}{-}   &0.027(0.021)  \\		
		\hline 		
		1000&naive lasso&4.00(0.00)&1.79(1.71)&1.00(0.00)&0.147(0.046)\\		
		~&PGEE.indep&4.00(0.00) &2.96(2.51)  & 1.00(0.00)  &0.034(0.028) \\
		~&PGEE.exch&4.00(0.00)&3.24(2.38) &1.00(0.00)  &0.018(0.014)\\
		~&PGEE.ar1&4.00(0.00)&3.28(2.29) &1.00(0.00)  &0.020(0.016)  \\		
		~&PWGEE.indep&4.00(0.00) &0.00(0.00)  & 1.00(0.00)  &0.012(0.008)  \\
		~&PWGEE.exch&4.00(0.00)&0.05(0.22) &1.00(0.00)  &0.027(0.022)  \\
		~&PWGEE.ar1&4.00(0.00)&0.06(0.24) &1.00(0.00)  &0.027(0.022) \\		
		~&Oracle.GEE.indep&\multicolumn{1}{c}{-}&\multicolumn{1}{c}{-} &\multicolumn{1}{c}{-}   &0.010(0.007)  \\
		~&Oracle.GEE.exch&\multicolumn{1}{c}{-}&\multicolumn{1}{c}{-} &\multicolumn{1}{c}{-}   &0.006(0.004)  \\
		~&Oracle.GEE.ar1&\multicolumn{1}{c}{-}&\multicolumn{1}{c}{-} &\multicolumn{1}{c}{-}  &0.007(0.005)  \\		
		~&Oracle.WGEE.indep&\multicolumn{1}{c}{-}&\multicolumn{1}{c}{-} &\multicolumn{1}{c}{-}  &0.011(0.007)\\
		~&Oracle.WGEE.exch&\multicolumn{1}{c}{-}&\multicolumn{1}{c}{-} &\multicolumn{1}{c}{-}  &0.024(0.016)  \\
		~&Oracle.WGEE.ar1&\multicolumn{1}{c}{-}&\multicolumn{1}{c}{-} &\multicolumn{1}{c}{-}  &0.025(0.016)  \\		
		\hline
	\end{tabular}
\end{table}

\begin{table}[t]
		\renewcommand{\arraystretch}{0.8}
	\centering
	\caption{Model selection results for Poisson regression without ICS}
	\label{poisson,nv=0}
	\vspace{0.2cm}
	\begin{tabular}{lccccc}
		\hline
		\multicolumn{1}{c}{$p_n$}	&Approaches & \multicolumn{1}{c}{TP} &\multicolumn{1}{c}{FP} &\multicolumn{1}{c}{CR}& \multicolumn{1}{c}{MSE}  \\
		\hline 				
		500&naive lasso&4.00(0.00)&1.84(1.60)&1.00(0.00)&0.146(0.059)\\		
		~&PGEE.indep&4.00(0.00) &2.88(2.12)  & 1.00(0.00)  &0.010(0.009)  \\
		~&PGEE.exch&4.00(0.00)&3.01(2.04) &1.00(0.00)  &0.007(0.006)\\
		~&PGEE.ar1&4.00(0.00)&2.97(2.02) &1.00(0.00)  &0.007(0.006) \\		
		~&PWGEE.indep&4.00(0.00) &0.00(0.00)  & 1.00(0.00)  &0.005(0.005) \\
		~&PWGEE.exch&4.00(0.00)&0.00(0.00) &1.00(0.00)  &0.010(0.009) \\
		~&PWGEE.ar1&4.00(0.00)&0.00(0.00) &1.00(0.00)  &0.010(0.009) \\		
		~&Oracle.GEE.indep&\multicolumn{1}{c}{-}&\multicolumn{1}{c}{-} &\multicolumn{1}{c}{-}  &0.003(0.003)  \\
		~&Oracle.GEE.exch&\multicolumn{1}{c}{-}&\multicolumn{1}{c}{-} &\multicolumn{1}{c}{-}  &0.002(0.002) \\
		~&Oracle.GEE.ar1&\multicolumn{1}{c}{-}&\multicolumn{1}{c}{-} &\multicolumn{1}{c}{-}  &0.003(0.002) \\		
		~&Oracle.WGEE.indep&\multicolumn{1}{c}{-}&\multicolumn{1}{c}{-} &\multicolumn{1}{c}{-}  &0.004(0.003)  \\
		~&Oracle.WGEE.exch&\multicolumn{1}{c}{-}&\multicolumn{1}{c}{-} &\multicolumn{1}{c}{-}  &0.009(0.007) \\
		~&Oracle.WGEE.ar1&\multicolumn{1}{c}{-}&\multicolumn{1}{c}{-} &\multicolumn{1}{c}{-}  &0.009(0.007)  \\		
		\hline
		1000&naive lasso&4.00(0.00)&1.96(1.91)&1.00(0.00)&0.173(0.071)\\		
		~&PGEE.indep&4.00(0.00) &3.27(2.12)  & 1.00(0.00)  &0.012(0.011) \\
		~&PGEE.exch&4.00(0.00) &3.21(1.98)  & 1.00(0.00)  &0.008(0.007) \\
		~&PGEE.ar1&4.00(0.00) &3.20(2.00)  & 1.00(0.00)  &0.009(0.007)   \\		
		~&PWGEE.indep&4.00(0.00) &0.00(0.00)  & 1.00(0.00)  &0.006(0.005)  \\
		~&PWGEE.exch&4.00(0.00)&0.00(0.00) &1.00(0.00)  &0.010(0.009) \\
		~&PWGEE.ar1&4.00(0.00)&0.00(0.00) &1.00(0.00)  &0.011(0.009) \\		
		~&Oracle.GEE.indep&\multicolumn{1}{c}{-}&\multicolumn{1}{c}{-} &\multicolumn{1}{c}{-}  &0.004(0.003)  \\
		~&Oracle.GEE.exch&\multicolumn{1}{c}{-}&\multicolumn{1}{c}{-} &\multicolumn{1}{c}{-}  &0.003(0.002) \\
		~&Oracle.GEE.ar1&\multicolumn{1}{c}{-}&\multicolumn{1}{c}{-} &\multicolumn{1}{c}{-}  &0.003(0.003)  \\
		
		~&Oracle.WGEE.indep&\multicolumn{1}{c}{-}&\multicolumn{1}{c}{-} &\multicolumn{1}{c}{-}  &0.004(0.003)  \\
		~&Oracle.WGEE.exch&\multicolumn{1}{c}{-}&\multicolumn{1}{c}{-} &\multicolumn{1}{c}{-}  &0.008(0.006)  \\
		~&Oracle.WGEE.ar1&\multicolumn{1}{c}{-}&\multicolumn{1}{c}{-} &\multicolumn{1}{c}{-}  &0.008(0.006) \\		
		\hline
	\end{tabular}
\end{table}

The selection and estimation results for the linear model and Poisson regression, without the influence of ICS, are summarized in Table~\ref{linear,nv=0} and Table~\ref{poisson,nv=0}, respectively. In the high-dimensional setting, the proposed PWGEE successfully identifies all important covariates, yielding a sparse model  with nearly zero FPs.  Compared with PGEE under correctly specified working correlation structure, the proposed PWGEE achieves comparable MSE and lower FP values. Naive lasso shows biased estimates. The estimation accuracy of PWGEE is nearly identical to that of the oracle WGEE, demonstrating their robustness to high-dimensional settings. By contrast, PGEE exhibits greater sensitivity to high-dimensional data.

In conclusion, the proposed PWGEE offers consistent model selection and robust estimation for analyzing high-dimensional longitudinal data, regardless of whether the outcomes are related to cluster size.

\section{Real Data Analysis}\label{Sec5}

Proteins are vital players in life processes, facilitating various physiological and biochemical reactions in the body. SARS-CoV-2 infection alters plasma proteome concentrations, which can lead to tissue damage, including lung inflammation, heart disease and long COVID \citep{Filbin2021, Haljasmagi2020, Shen2020, Yin2024}. Researchers are keen to uncover the regulatory mechanisms of plasma proteomes linked to COVID-19 severity, aiming to provide effective treatments for patients.

The plasma proteomic data includes 384 patients suspected to have COVID-19, along with several of their physical signs over a 28-day period following enrollment \citep{Filbin2021}. For our analysis, we focus on observations from  $n=305$ COVID-19 positive patients at Day 0, as well as those who remain hospitalized on Day 3 and Day 7, yielding a total of $N=659$ measurements. A high dimensional set of $p_n=1,472$ plasma proteins is measured using the proximity extension assay by Olink Proteomics. Protein levels are expressed as normalized protein expression value on a log2 scale. We treat disease severity, $Y_{ij}$, as a binary outcome. Following \cite{Filbin2021}, we classify acuity levels into A1–A5 on Days 0, 3 and 7, based on the World Health Organization  guidelines.  Severe cases ($Y_{ij}=1$) include A1 (deceased) and A2 (intubated, survived), whereas non-severe cases ($Y_{ij}=0$) encompass  outcomes A3 (hospitalized on oxygen),
A4 (hospitalized without oxygen), and A5 (discharged from the Emergency Department).

\begin{table}
		\renewcommand{\arraystretch}{0.8}
	%	\centering
	\caption{Selected model sizes for identifying the proteins in COVID-19 pathological process.} \label{sel_pro}
	\vspace{0.2cm}
	\setlength{\tabcolsep}{4.5mm}{
		\begin{tabular}{lccccccc}
			\hline
			\multirow{2}{*}{Approaches} & \multirow{2}{*}{naive lasso}	 &
			\multicolumn{3}{c}{PGEE}& 	\multicolumn{3}{c}{PWGEE}\\ \cmidrule(r){3-5}\cmidrule(r){6-8}
			& & indep  & exch&
			ar1& 	indep& exch& ar1\\ \hline
			\multicolumn{1}{c}{Model size}
			&26 & 14 & 30 & 26 &16 &16 &15\\
			\hline
		\end{tabular}	
	}
\end{table}

We apply the proposed PWGEE to analyze the plasma proteomic data.
To identify proteins involved in the pathological process of COVID-19, we model the marginal mean of $Y_{ij}$ as
\begin{equation}
	\log\frac{\mu_{ij}}{1-\mu_{ij}}=\beta _0+\sum_{d=1}^{p_n}\beta_{d}x_{ij,d},\label{COV-model}
\end{equation}
where $Y_{ij}$  represents the COVID-19 severity for patient $i$ measured at time point $j$ for $i=1,\dots, 305$ and $j=1,\dots,3$, and $x_{ij,d}$ is the normalized protein expression value of the $d$-th protein for the $i$-th positive patient at time point $j$. We apply a penalty to the magnitude of the $\beta_d$ coefficients  for $d=1,\dots,1,472$.

\begin{figure}[t]
	\vspace{2cm}
	\begin{center}

		\includegraphics[width=5in, height=5in]{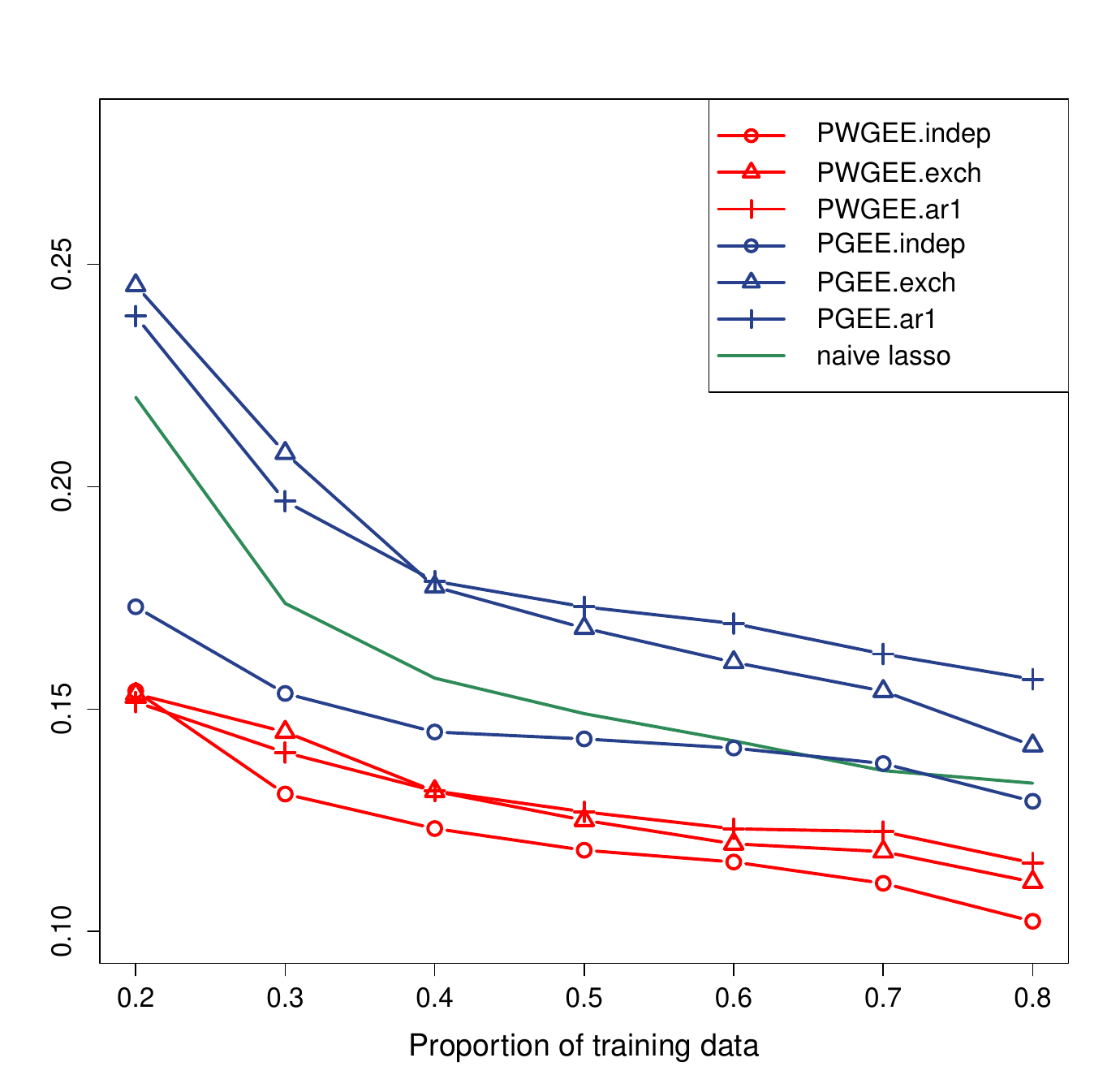}
	\end{center}
	\vspace{-0.5cm}
	\caption{Median test classification errors using various proportions of data as training sets for different model selection methods.\label{class_plot}}	
\end{figure}

We compare the performance of our PWGEE with naive lasso \citep{Tibshirani1996} and PGEE \citep{Wang2012} under independent,  exchangeable and autoregressive working correlation structures. Table~\ref{sel_pro} presents the model selection sizes for each method.
Among them, the proposed PWGEE consistently identified moderate and robust model sizes across all three correlation structures. Several proteins identified by PWGEE were confirmed to be closely associated with COVID-19 severity.
For example, PLA2G2A plays a crucial role in cell death, lipid metabolism and the secretion of inflammatory factors. When its plasma concentration exceeds 10 ng/mL, the mortality risk for patients with COVID-19  significantly increases \citep{Snider2021}.
Similarly, protein EZR was elevated in patients with severe neuro-COVID,  compared to those with other central nervous system inflammatory conditions, highlighting its potential as a predictive biomarker \citep{Etter2022}.

We also assess the classification errors of the proposed PWGEE with competing methods. To show the impact of data size, we vary the proportion of training data from 20\% to 80\% in increments of 10\% and use the remaining data for testing. Each split is repeated 100 times, and we report the median classification errors.  Figure~\ref{class_plot} summarizes the results.
The proposed PWGEE method almost always outperforms the other methods in accurately predicting whether a patient's condition is severe or not.
Combined with the results in Table~\ref{sel_pro}, we conclude that other methods, particularly PGEE.exch and PGEE.ar1, exhibit varying degrees of overfitting.
When the training data proportion is 20\%, the classification accuracy of  PWGEE  exceeds 80\% and approaches 90\% when 80\% of the patients are used for training.  Besides, naive lasso shows higher classification errors than the proposed PWGEE as it ignores within cluster correlations and relies  on the independence among predictors.
Although PGEE shows decreasing classification errors as the training data proportion increases, it performs worse under exchangeable and autoregressive working correlation structures, indicating severe overfitting and the selection of redundant proteins.

In conclusion, the proposed PWGEE method outperforms the  three  other methods in identifying proteins involved in the pathological development following COVID-19 infection.

\section{Discussions}\label{Sec6}

The GEE method is widely used in longitudinal data analysis. However, the  GEE method results in biased estimates  when the outcome of interest is associated with cluster size, a phenomenon known as ICS. In this article, we introduce  a novel WGEE approach to eliminate the  impact of ICS and extend its penalized version (PWGEE) to high-dimensional longitudinal data. Through rigorous theoretical analysis and validation with simulation examples and real data analysis, we found that the proposed PWGEE offers consistent model selection and robust estimation for analyzing high-dimensional longitudinal data, regardless of whether the outcomes are related to cluster size. Therefore, the WGEE and PWGEE methods proposed in this article offer a unified framework for robust longitudinal data analysis, based on assumptions regarding the first and second moments of the data.

\end{document}